\documentclass[review]{elsarticle}

\usepackage{lineno,hyperref}
\usepackage{graphicx}
\usepackage[letterpaper]{geometry}
\usepackage{amsmath}
\usepackage{amssymb}
\usepackage{bm}

\modulolinenumbers[5]

\journal{Journal of \LaTeX\ Templates}

\begin{document}

\begin{frontmatter}

\title{Hyperspherical Slater determinant approach to few-body fractional quantum Hall states}

\author{Bin Yan, Rachel E. Wooten, Kevin M. Daily, and Chris H. Greene}
\address{Department of Physics and Astronomy, West Lafayette, IN}




\begin{abstract}
In a recent study\cite{Daily:2015}, a hyperspherical approach has been developed to study of few-body fractional quantum Hall states. This method has been successfully applied to the exploration of few boson and fermion problems in the quantum Hall region, as well as the study of inter-Landau level collective excitations\cite{Rittenhouse:2016,Wooten:2016}. However, the hyperspherical method as it is normally implemented requires a subsidiary (anti-)symmetrization process, which limits its computational effectiveness. The present work overcomes these difficulties and extends the power of this method by implementing a representation of the hyperspherical many-body basis space in terms of Slater determinants of single particle eigenfunctions. A clear connection between the hyperspherical representation and the conventional single particle picture is presented, along with a compact operator representation of the theoretical framework.
\end{abstract}


\end{frontmatter}


\section{Introduction}\label{sec_intro}
In a strong magnetic field, electrons condense into phases that are often described consisting of fractionally charged quasi-particles\cite{Clark1988,Mahalu1997} that can obey fractional statistics. These new phases are the well-known quantum Hall states\cite{Klitzing:1980,Tsui:1982,Laughlin:1983} which cannot be classified within Landau's symmetry breaking picture. In typical experiments on two-dimensional electron systems, the Hall resistance and the magnetoresistance are shown to have quantized values at integer and certain fractional values of the filling factor, which is the ratio between the number of electrons and the degeneracy of single particle states in a finite area.

Extensive headway has been made in developing a theoretical picture of this effect\cite{Jain:1989,Haldane:1983,Laughlin:1983pb,Haldane:1985,Halperin:1984,MooreRead:1991,ReadRezayi:1999,Girvin:1987,Bernevig:2009dn,Jain:2001vg,Halperin:1982ej}, but it still remains far from being completely understood, especially in cases where higher Landau levels are involved.
Continued interest in reproducing and studying the fermionic quantum Hall effect and its bosonic analog in highly-controlled atomic systems also demands innovative and unconventional methods of studying the few-body quantum Hall effect. In a recent line of attack\cite{Daily:2015}, a novel approach to the quantum Hall problem was presented that is based on the adiabatic hyperspherical representation\cite{Macek:1968,Greene2010,Greene2011}, which originated in and has been extensively used in the context of few-body physics\cite{Greene2011,Macek1999,Viviani2012,Lin1995}. The hyperspherical approach not only provides complementary advantages and alternative qualitative pictures compared to previous methods, it is also more suitable for the discussion of few-body systems (i.e., cold atoms in rotating traps, electrons in a quantum dot\cite{Jolicoeur2003,Dalibard2001,Lewenstein2005,Cornell2004,Lukin2005}). Other recent studies have successfully applied the hyperspherical method to the study of the two-dimensional three-boson problems\cite{Rittenhouse:2016} in the presence of a perpendicular magnetic field, and to the study of inter-Landau level collective excitations\cite{Wooten:2016}.

The quantum Hall effect is conventionally understood starting from the single particle representation, where the many-body wave functions are constructed directly from single particle basis functions. (Details of this are summarized in Section \ref{sec_single}.) The hyperspherical approach, however, tackles the problem from a collective perspective. Through a coordinate transformation, this method solves the many-body Schr\"odinger equation directly, the eigensolutions of which are given in terms of hyperspherical harmonics, functions of a set of hyperangles in these collective coordinates. (See Section \ref{sec_hyperspher} for details.) This point of view highlights many key properties of the system which do not emerge naturally from the independent particle framework. However, a drawback of the hyperspherical method is that the hyperspherical harmonics do not {\it a priori} possess any intrinsic particle permutation symmetry, while the many-fermion (or many-boson) wavefunctions are required to be antisymmetric (symmetric) under particle exchange. As a consequence, an antisymmetrization (symmetrization) process should normally be performed in order to generate basis functions with the proper exchange symmetry\cite{Daily:2015,Efros:1995,Barnea:1999,Krivec:1998,Vivian:1998}. (An alternative method sometimes used to attack this symmetrization problem is via postsymmetrization\cite{Viviani2012}, which we do not pursue here, because we prefer to work with far smaller pre-symmetrized basis sets.) This symmetrization step is the main bottleneck limiting the computational power of the hyperspherical method. In order to solve this problem, we implement here a basis of Slater determinants (permanents) which can overcome this difficulty because they are explicitly antisymmetrized (symmetrized). We establish here the reduction of the conventional Slater determinant basis functions to the many-body hyperspherical basis function spaces. A similar technique has previously been implemented in some nuclear physics calculations\cite{Fabre:1978,Fabre:2005,Timofeyuk:2002,Timofeyok:2004}. This reduction allows us to extend the hyperspherical method to 8 electrons in the $1/3$ fractional quantum Hall region, with computational expenses comparable to our previous method implemented in Ref.\cite{Daily:2015}, which could treat at most 6 electrons. 

This paper is organized as follows. Section \ref{sec_single} defines the quantum Hall problem and describes the single particle state space picture of the conventional Slater determinant(permanent) approach. Section \ref{sec_hyperspher} introduces the hyperspherical approach to the quantum Hall problem. Section \ref{sec_representation} offers the prescription for generating all the Slater determinant basis functions that can contribute to a given hyperspherical $\{K, M\}$ manifold, and then derives the form of hyperspherical operators in terms of single particle ladder operators. Section \ref{sec_summary} gives some concluding remarks. Finally, the Appendix offers the detailed derivations of the relevant operators, including the two-body interaction matrix elements. 
 
\section{Single particle representation}\label{sec_single}
\subsection{Single particle Hamiltonian}
The Hamiltonian for a single electron in two dimensions with a perpendicular magnetic field is given by
\begin{equation}\label{eq_singleH}
H=\frac{1}{2m_e}(-\imath\hbar\bm{\nabla}+e\bm{A})^2,
\end{equation}
where $m_e$ is the electron mass, $e$ is the magnitude of the electron charge, $\hbar$ is Planck's constant, and $\bm{A}$ is the vector potential. The symmetric gauge, $\bm{A}= (B/2)(-y \hat{x}+ x \hat{y})$, is used here to represent the constant magnetic field of magnitude $B$
oriented in the positive $\hat{z}$ direction. With this choice of $\bm{A}$, Eq. (\ref{eq_singleH}) can be written in Cartesian coordinates as
\begin{equation}
  \label{eq_Hsingle}
  H = - \frac{\hbar^2}{2m_e} \bm{\nabla}^2 
  + \frac{e^2 B^2}{8m_e }(x^2+y^2)
  + \frac{e B}{2m_e} L_{z}, 
\end{equation}
where $L_{z} = -\imath\hbar(x \partial_{y} - y \partial_{x})$ is the $z$-component of the angular momentum operator.
In the magnetic length scale where length is expressed in the units of $\lambda_0$,
\begin{equation}
  \label{eq_maglength}
  \lambda_0 = \sqrt{\frac{\hbar}{m_e \omega_c}},
\end{equation}
the Hamiltonian (\ref{eq_singleH}) takes the following form in polar coordinates: 

\begin{equation}
  \label{eq_H1r}
  H = -\frac{1}{2} \left\{ \frac{1}{r} \partial_r 
  r \partial_r 
  - \frac{L_z^2}{\hbar^2 r^2}\right\}
  + \frac{1}{8} r^2
  + \frac{1}{2\hbar} L_{z}.
\end{equation}
The solutions to the Schr\"odinger equation with this Hamiltonian are
\begin{equation}
  \Psi_n^m(\bm{r}) = \sqrt{\frac{n!}{\Gamma(n+|m|+1) 2^{|m|}}}
    e^{-r^2/4} L_n^{(|m|)}(r^2/2) r^{|m|} e^{\imath m \phi},
\end{equation}
with single-particle eigenenergies
\begin{align}
  \label{eq_NIenergy}
  E^{(1)} = \frac{1}{2} \left(2 n + m + |m| + 1 \right).
\end{align}
Here the energy is expressed in units of $\hbar \omega_c$, where $\omega_c \equiv eB/m_e$ is the cyclotron frequency. $n$ and $m$ are the radial quantum number and the rotational quantum number about the z-axis respectively. 
It is more convenient to use the Landau level label $\epsilon$ instead of the radial quantum number $n$; these are related by \begin{equation}
  \epsilon = n + \frac{m+|m|}{2}.
\end{equation}
States with the same Landau level label $\epsilon$ all share the same energy
\begin{equation}
  \label{eq_NIenergy2}
  E^{(1)} = \epsilon + \frac{1}{2}.
\end{equation}

\subsection{Ladder operators and Landau levels}
The single-particle states are expressed in bra-ket notation, $|\epsilon,m\rangle$,
and the well-known ladder operators cause transitions among the single-particle states\cite{Jain:2007}.
The effect of the operators on the kets is summarized below:
\begin{equation}
\begin{aligned}
  b \; |\epsilon, m\rangle &=\left\{
  \begin{array}{rc}
    +1, & \mbox{ $m < 0 $} \\
    -1, & \mbox{ $m \ge 0$}
  \end{array}
  \right\} \;\sqrt{\epsilon-m}\;|\epsilon, m+1\rangle \\
  b^{\dagger} \; |\epsilon, m\rangle &=\left\{
  \begin{array}{rc}
    +1, & \mbox{ $m \le 0 $} \\
    -1, & \mbox{ $m > 0$}
  \end{array}
  \right\} \;\sqrt{\epsilon-m+1}\;|\epsilon, m-1\rangle \\
  a \; |\epsilon, m\rangle &=\left\{
  \begin{array}{rc}
    +1, & \mbox{ $m > 0 $} \\
    -1, & \mbox{ $m \le 0$}
  \end{array}
  \right\}  \;\sqrt{\epsilon}\;|\epsilon-1, m-1\rangle \\
  a^{\dagger} \; |\epsilon, m\rangle &=\left\{
  \begin{array}{rc}
    +1, & \mbox{ $m \ge 0 $} \\
    -1, & \mbox{ $m < 0$}
  \end{array}
  \right\}  \;\sqrt{\epsilon+1}\;|\epsilon+1, m+1\rangle \\
  ab \; |\epsilon, m\rangle &= -\sqrt{\epsilon}\sqrt{\epsilon-m} \; |\epsilon-1,m\rangle \\
  a^{\dagger}b^{\dagger} \; |\epsilon, m\rangle &=
  -\sqrt{\epsilon+1}\sqrt{\epsilon+1-m} \; |\epsilon+1,m\rangle
\end{aligned}
\end{equation}

The lowest rung of each ladder obeys
\begin{equation}
\begin{aligned}
  b \; |\epsilon, \epsilon \rangle &= 0 \\
  a \; |0, m\rangle &= 0.
\end{aligned}
\end{equation}

\begin{figure}
  \centering
  \includegraphics[angle=0,width=0.75\textwidth]{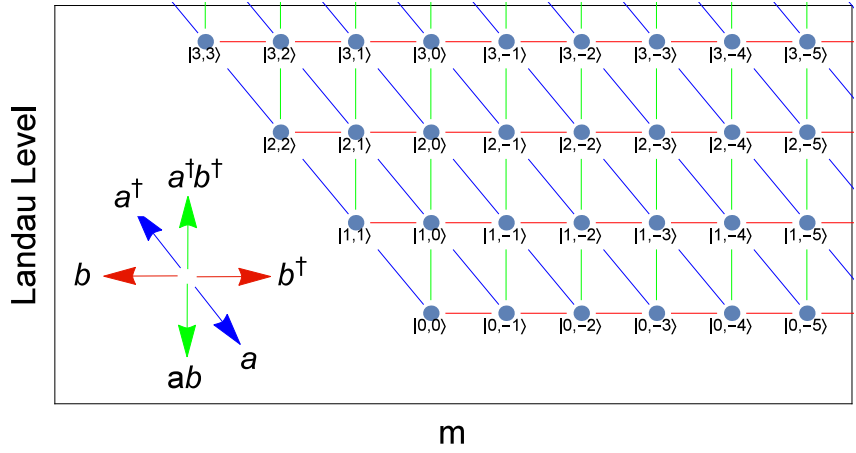} 
  \caption{(Color online) 
    Diagram describing the Landau levels and how the ladder
    operators move among the states $|\epsilon,m\rangle$,
    where $\epsilon$ is the Landau level and $m$ is the 2-D
    angular momentum quantum number.
  }
  \label{fig_LL}
\end{figure}

Figure \ref{fig_LL} depicts the single-particle states,
arranged according to $m$ along the horizontal
and Landau level $\epsilon$ along the vertical.
Red lines indicate states that are neighbors according to
the $b$ and $b^{\dagger}$ operators.
The $b$ operators raise the $m$ quantum number up to
the maximum value where $m$ is equal to the Landau level,
e.g. for the lowest level $m_{\rm max}=0$.
Blue lines indicate states that are neighbors according to
the $a$ and $a^{\dagger}$ operators.
The $a$ operators lower the Landau level
to a minimum value of $\epsilon=0$,
while simultaneously modifying the $m$ quantum number
by one unit each time $a$ or $a^{\dagger}$ is applied.
Green lines indicate states that are neighbors according to
the $ab$ and $a^{\dagger}b^{\dagger}$ operators.
These combined operators raise or lower the Landau level
while preserving the $m$ quantum number.

\subsection{Many-body state space}
With the single particle basis function in hand, the many-body basis functions can be constructed as Slater determinants or permanents, depending on the underlying particle statistics. Problems that include interactions can then be subsequently solved by expanding into this many-body basis set. It is worth noting that because typical interactions that are pairwise and central (e.g., Coulomb interaction and p-wave contact interaction), there are no couplings between Slater determinants or permanents with different total angular momenta ($M_{tot}=\sum_{i}m_i$). In other words, $M_{tot}$ is a good quantum number. Thus one can talk about the Hilbert subspace $\{M_{tot}\}$ of a given total angular momentum. One can further separate out the center-of-mass motion and define the fixed relative angular momentum subspace $\{M_{rel}\}$ as a subspace of $\{M_{tot}\}$, in which all basis functions have vanishing center-of-mass angular momentum. 
In the following hyperspherical approach, we also define another quantum number, $K$, exact for the noninteraction system and still an approximately good quantum number in the presence of interactions, which labels the eigenvalue of the grand angular momentum operator $\hat{K}^2$. The resulting subspace with fixed values of $K$ and $M$ is then called the $\{K,M\}$ manifold. (In the following, $M$ always refers to $M_{rel}$.)

\section{Hyperspherical representation}\label{sec_hyperspher}

\subsection{Relative Hamiltonian}
In contrast to the conventional approach, where the many-body states are constructed using single particle basis, the hyperspherical method treats the many-body Hamiltonian collectively, namely, the many-body basis functions are solved directly from the Schr\"odinger equation. To achieve this, we first separate the N-body noninteracting Hamiltonian $H_N$
into center-of-mass ($H_{CM}$) and relative ($H_{rel}$) components. The center-of-mass behaves like an independent single particle, while the non-interacting relative Hamiltonian takes the form of

\begin{equation}
  \label{eq_HN}
\begin{aligned}
  H_{\rm rel} = &- \frac{1}{2\mu}\sum_{j=1}^{N_{\rm rel}} \bm{\nabla}_j^2 
  + \frac{\mu}{8}\sum_{j=1}^{N_{\rm rel}} (x_j^2 + y_j^2) 
  + \frac{1}{2\hbar}\sum_{j=1}^{N_{\rm rel}} L_{z_j}^{\rm rel},
\end{aligned}
\end{equation}
where $x_j$ and $y_j$ are the Cartesian components of $N_{\rm rel}=N-1$ relative Jacobi vectors $\bm{\rho}_j$, and $\mu$ is a dimensionless mass scaling factor\cite{Delves:1959,Delves:1960},
\begin{align}
  \label{eq_mu}
  \mu = \left( \frac{1}{N} \right)^{1/N_{\rm rel}}.
\end{align}
The definition of Jacobi vectors in terms of single particle coordinates is arbitrary. (As an example of the transformation, see Sec. III in Ref.\cite{Daily:2015})

\subsection{Hyperspherical transformation}

The hyperspherical coordinates are a high dimension analogue of the three-dimensional spherical coordinates. The overall size of the system is characterized by a single scalar coordinate, the hyperradius R, which is defined as 
\begin{equation}
  \label{eq_hyperradius}
  R^2=\sum_{j=1}^{N_{\rm rel}} \rho^2_j.
\end{equation}

The remaining degrees of freedom, which represent the geometry of the system, are encoded in a set of coordinates, the hyperangles $\bm{\Omega}$. The definition of the hyperangles has some arbitrariness, and there are many different schemes in the literature\cite{Smirnov:1977,Avery:1989,Avery:1993,Rittenhouse:2006}. The major focus of this paper does not depend on the specific definition of the hyperangles. To see a concrete example of this transformation, refer to Sec. IV.A in Ref\cite{Daily:2015}.

\subsection{Hyperspherical basis functions and the grand angular momentum operator}

Under the transformation to hyperspherical coordinates, the relative non-interacting Hamiltonian, Eq.~\eqref{eq_HN}, transforms to
\begin{equation}
  \label{eq_HNhs}
  H_{\rm rel} = - \frac{1}{2\mu} \bm{\nabla}^2_{R, {\bm \Omega}} 
  + \frac{\mu}{8} R^2 
  + \frac{1}{2\hbar}L_z^{\rm rel,tot}.
\end{equation}
Here $L_z^{\rm rel,tot}$ is the $z$-component
of the total relative angular momentum. The Laplacian operator in hyperspherical coordinates is given by
\begin{align}
  \bm{\nabla}^2_{R, {\bm \Omega}} =   \frac{1}{R^{2N_{\rm rel}-1}} \partial_R
    R^{2N_{\rm rel}-1} \partial_R
    - \frac{\hat{\bm{K}}^2}{R^2}.
\end{align}
$\hat{\bm{K}}$ is called the grand angular momentum operator\cite{Avery:1989}, whose eigenfunctions are represented by the orthonormal hyperspherical harmonics $\Phi_{K{u}}^{(M)}(\bm{\Omega})$, where
\begin{equation}
\hat{\bm{K}}^2 \Phi_{K{u}}^{(M)}(\bm{\Omega}) 
   = K (K + 2N_{\rm rel} - 2)\Phi_{K{u}}^{(M)}(\bm{\Omega}).
\end{equation}
Here, the subscript $u$ labels different degenerate eigenfunctions with the same $K$ and $M$. $M$ is the quantum number of the total relative angular momentum, which is a good quantum number for any central two-body potential. The hyperspherical harmonics $\Phi_{K{u}}^{(M)}(\bm{\Omega})$ are the simultaneous eigenstates of $\hat{\bm{K}}^2$ and $L_z^{\rm rel,tot}$ under constraint $|M| \le K$.

The full eigensolutions of non-interacting relative Hamiltonian (\ref{eq_HNhs}) are separable into post-(anti)symmetrized hyperspherical harmonics $\Phi_{Ka}^{(M)}(\bm{\Omega})$ and hyperradial functions $F_{n_RK}^{(M)}(R)$,
\begin{align}
  \label{eq_wf}
  \Psi(R,\bm{\Omega}) = R^{-N_{\rm rel}+1/2}
   F_{n_RK}^{(M)}(R)\Phi_{K{a}}^{(M)}(\bm{\Omega}).
\end{align}
The hyperradial functions $F_{n_RK}^{(M)}(R)$ satisfy a one-dimensional differential equation:
\begin{align}
  \label{eq_HNI}
  & \left\{ -\frac{1}{2\mu} \frac{d^2}{dR^2}
  + U_{K}^{(M)}(R) - E \right\} F_{n_RK}^{(M)}(R)
  = 0,
\end{align}
where the hyperradial potentials $U_{K}^{(M)}(R)$
are given by
\begin{align}
  \label{eq_UNI}
  U_{K}^{(M)}&(R) =  \\ \nonumber
  &\frac{(K+N_{\rm rel}-1/2)(K+N_{\rm rel}-3/2)}{2\mu R^2}
  + \frac{\mu}{8}R^2 + \frac{1}{2}M.
\end{align}
 
The many-body states (\ref{eq_wf}) serve as the basis for the study of interacting problems. In practice, an adiabatic approximation that initially treats the hyperradius $R$ as a parameter is adapted, in other words, the fixed $R$ Hamiltonian is diagonalized at a fixed hyperradius in the space of hyperspherical harmonics.

We note the following remarks. 1) The forms of the hyperspherical harmonics $\Phi_{K{u}}^{(M)}(\bm{\Omega})$ depend on the coordinate transformation scheme. They are simple functions to express in hyperspherical coordinates, but become cumbersome to use as the computational basis when re-expressed in particle Cartesian coordinates. 2) The grand angular momentum quantum number $K$ ($K=0,1,2,\ldots$) is a good quantum number in the noninteracting limit, and it has been demonstrated to be an approximate good quantum number even in the presence of Coulomb interactions. 3) There are $(2N_{\rm rel}-2+2K)(2N_{\rm rel}-3+K)!/(K !(2N_{\rm rel}-2)!)$ linearly independent hyperspherical harmonics in a given $K$ manifold. However, these basis functions generally do not possess the proper symmetry under particle permutation. Thus a process of finding all linear combinations that have the desired symmetric must be performed, which makes the use of the hyperspherical basis functions challenging. The following develops a method to minimize the above difficulties, by implementing a representation of the anti-symmetrized hyperspherical harmonics in terms of Slater determinant of single particle basis functions (See also Refs.\cite{Rittenhouse:2006,Fabre:2005,Fabre:1978,Timofeyuk:2002,Timofeyok:2004}).

\section{Representation of the $\{K, M\}$ manifold harmonics in terms of the Slater determinant basis}\label{sec_representation}

The purpose of this section is to reduce a given hyperspherical $\{K, M\}$ manifold to a single particle representation, namely the space spanned by Slater determinants or permanents of single particle basis functions. 

\subsection{Enumerating the Slater determinants of a $\{K, M_{tot}\}$  manifold}\label{sec_enumerating}

The first task is to find all possible Slater determinants
that can form part of a given $\{K, M_{tot}\}$ manifold with the center of mass included.
Here,
$M_{tot}$ is the total projection quantum number, $M_{tot} = \sum_{j=1}^N m_j$,
and $K$ is the grand angular momentum quantum number.
$K$, as was shown in a previous study\citep{Daily:2015},
is an approximately good quantum number in the many-body
quantum Hall problem.
In fact, $K$ is equal to the order of the harmonic polynomial
of the many-body wave function in a fixed $\{K, M_{tot}\}$ manifold. 
Producing the list of $N$-particle Slater determinants that span a given fixed $\{K, M_{tot}\}$ manifold is equivalent to finding the complete list of sets of $N$ single-particle orbitals that satisfy the obey a short list of restrictions:
\begin{enumerate}
\item All $N$ single particle orbitals selected for a given Slater determinant must be allowed:
\begin{itemize}
\item The maximum order of the polynomial part of any single particle orbital, given by $k_i = 2\epsilon_i-m_i$ for the i$^{th}$ orbital must be greater than or equal to zero,
$2\epsilon_i -m_i \geq 0$
\item The radial quantum number, $n_i$, for the i$^{th}$ orbital must be non-negative, $n_i \geq 0$. This restricts the selection of $\epsilon_i = n_i+(m_i+|m_i|)/2$,
\end{itemize}
\item The total angular momentum $M_{tot}$ is the sum of the single-particle $m_i$ values,
\begin{equation}
M_{tot} = \sum_{i=1}^N m_i.
\end{equation} 
\item The total grand angular momentum, $K$, is equal to the total order of the N-particle harmonic polynomial of the product of the selected $N$ single-particle orbitals. This is determined by the rule
\begin{equation}
K = \sum_{i=1}^N(2\epsilon_i - m_i).
\end{equation}
The total orders of the single particle polynomials are represented graphically in Fig. \ref{fig_contingency}.
\item For fermions, the orbitals must all be different in order to satisfy the Pauli exclusion principle.
\end{enumerate}
Each set of $N$ single-particle orbitals that satisfies this list of rules defines a single Slater determinant in the desired $\{K, M_{tot}\}$ manifold.  The complete set of all Slater determinants that satisfy these rules spans the entire $\{K, M_{tot}\}$ manifold basis, and any totally hyperspherical function in that $\{K, M_{tot}\}$ manifold that is antisymmetric with respect to particle-interchange can be expressed as a linear combination of these Slater determinants.   

This Slater determinant list can be found directly by testing all single-particle orbital sets that satisfy this list of rules, although the procedure is somewhat tedious to carry out by hand and requires significant testing. 
An alternative streamlined and more systematic method of finding the complete set of Slater determinants of a fixed $\{K,M_{tot}\}$ manifold using integer partitions and contingency tables is described in \ref{appendixKM}.

\subsection{Operator diagonalizations}\label{sec_diag}
If the functional space is reduced to states
in which the center of mass is in its absolute ground state,
that is, there are no Landau level or rotational excitations in the center of mass,
then that also implies that $K_{\rm CM}=0$.
Thus,
finding the eigenvalues of $K_{\rm tot}$
for the set of center-of-mass-reduced states
is effectively equivalent to the eigenvalues of $K_{\rm rel}$.

The relevant operators for this work are
\begin{align}
  L_{CM} &= \frac{\hbar}{N} \sum_{j=1}^N
  \left( a^{\dagger}a- b^{\dagger}b \right)_j
  + \frac{\hbar}{2N} \sum_{j=1}^N \sum_{k \ne j}^N
  \left[a_k^{\dagger}a_j + a_j^{\dagger}a_k
    - \left(b_k^{\dagger}b_j +b_j^{\dagger}b_k\right)  \right],
  \\
  H_{CM} &= \frac{1}{2} + \frac{1}{N} \sum_{j=1}^N a_j^{\dagger} a_j
  + \frac{1}{2N} \sum_{j=1}^N \sum_{k \ne j}^N
  \left( a_k^{\dagger}a_j + a_j^{\dagger}a_k  \right),
\end{align}
and
\begin{align}
  \hat{K}^2 &=  \sum_{j=1}^N
  \left(
  \left[a^{\dagger}a - b^{\dagger}b\right]_j^2
  +(2N-2) \left[a^{\dagger}a+b^{\dagger}b\right]_j
  \right)
  \nonumber \\ &
  +\sum_j \sum_{k \ne j}
  \left(
  \left[a^{\dagger}a+b^{\dagger}b \right]_j\left[a^{\dagger}a+b^{\dagger}b\right]_k
  -2\left(a_j^{\dagger}b_j^{\dagger}a_kb_k+a_jb_ja_k^{\dagger}b_k^{\dagger} \right)
  \right).
\end{align}
The detailed derivations of these operator expressions are given in \ref{appendix_operator}.

In practice,
for a given set of Slater determinant basis functions,
the $L_{CM}$ operator is diagonalized and the eigenvectors corresponding to 0 eigenvalue are selected.
Next, the $H_{CM}$ operator is diagonalized in the basis of $L_{CM}=0$ states
and the eigenvectors with corresponding $1/2$ eigenvalues are selected.
Lastly, if necessary,
the $\hat{K}^2$ operator is diagonalized
in the basis of $L_{CM}=0$, $E_{CM}=1/2$,
and the eigenvectors with corresponding $K(K+2N-2)$ eigenvalues
(the minimum eigenvalues) are selected.
In the lowest Landau level,
there are zero Landau level excitations
and all states having $L_{CM}=0$ turn out to have the smallest possible value of $K(K+2N-2)$, which are the eigenvalues of the $\hat{K}^2$ operator.
Thus only the $L_{CM}$ operator needs to be diagonalized.

\subsection{Coulomb matrix elements at a fixed hyperradius}\label{sec_CoulombMatrix}
We assume we have basis functions $\Psi(\bm{r}_1,\bm{r}_2,\ldots)$
in the independent particle Slater determinant representation
with $L_{CM}=0$, $E_{CM}=1/2$, and fixed $M$.
We may also sometimes need matrix elements between basis functions having different values of $K$.
Matrix elements at a fixed hyperradius $R$ can be computed
by equating the integral over all Cartesian coordinates
with the integral over the center of mass and the relative function
expressed in hyperspherical coordinates.
Our starting point is
\begin{align}
  \label{eq_equate}
  &\int \Psi_{K'}^*(\bm{r}_1,\bm{r}_2,\ldots) V(\bm{r}_{12})
  \Psi_K(\bm{r}_1,\bm{r}_2,\ldots) d\bm{r}_1d\bm{r}_2\ldots
  \nonumber \\ &
  \qquad = \int |\Psi(\bm{R}_{CM})|^2d\bm{R}_{CM}
  \int \Psi_{K'}^*(R;\bm{\Omega}) V(R;\bm{\Omega}) \Psi_K(R;\bm{\Omega})
  R^{2N-1}dRd\bm{\Omega}
\end{align}
where it is assumed that for every basis function the center of mass
is in its absolute ground state and can be separated off.
The basis functions are labeled by the $K$ quantum number.
The left-hand-side of Eq.~\eqref{eq_equate} is assumed to be known
from standard Slater determinant basis methods and we label it $I$.
The center of mass integral on the right-hand-side of Eq.~\eqref{eq_equate}
is unity.
Moreover,
the hyperradial and hyperangular parts of the basis functions are known.
This leaves
\begin{align}
  \label{eq_equate}
  I &= \int \Psi_{K'}^*(R;\bm{\Omega}) V(R;\bm{\Omega}) \Psi_K(R;\bm{\Omega})
  R^{2N-1}dRd\bm{\Omega}
  \nonumber \\ &
  = \int \Phi_{K'}^*(\bm{\Omega}) \left(\int {\cal N}^* e^{-\tfrac{\mu}{4}R^2}R^{K'} 
  V(R;\bm{\Omega}) {\cal N} e^{-\tfrac{\mu}{4}R^2}R^K R^{2N-1}dR \right)
  \Phi_K(\bm{\Omega})   d\bm{\Omega}
\end{align}
where ${\cal N}$ is the normalization of the hyperradial wave function.
In the case of power law potentials,
the hyperradial dependence of the interaction potential $V$ is separable,
$V(R;\bm{\Omega})=R^pV(\bm{\Omega})$.
Thus the hyperradial integral can be factored
from the hyperangular integral, which leaves
\begin{align}
  \label{eq_equate}
  \frac{I}{\langle K'| R^p |K\rangle} &=
  \int \Phi_{K'}^*(\bm{\Omega}) V(\bm{\Omega}) \Phi_K(\bm{\Omega})d\bm{\Omega}
\end{align}
where $\langle K'| R^p |K\rangle$ is the hyperradial matrix element,
\begin{align}
  \langle K'| R^p |K\rangle =
  \left( \frac{2}{\mu} \right)^{p/2}
  \frac{\Gamma\left( [K+K'+2N+p]/2 \right)}
       {\sqrt{\Gamma\left(K+N\right)\Gamma\left(K'+N\right)}}.
\end{align}

\subsection{Two-body matrix elements of the Coulomb potential}

According to the Slater-Condon rules, the matrix elements of any two-body operators $\hat{O}$ in the basis of N-body Slater determinants $|\epsilon_1,m_1\rangle|\epsilon_2,m_2\rangle...|\epsilon_N,m_N\rangle$ can be expressed as a sum in terms of two-body matrix elements $\langle\epsilon 1,m1|\langle\epsilon2,m2|\hat{O}_{12}|\epsilon1',m1'\rangle|\epsilon2',m2'\rangle$, where the number $1$ and $2$ label the two particles. In the case of Coulomb interactions, or a general class of interactions where the potential depends only on the inter-particle distance $r$, it is more convenient to compute the two-body matrix element in terms of center-of-mass and relative coordinates, 
\begin{equation}
\langle N ,M|\langle n,m|\hat{O}(r)|N',M'\rangle|n',m'\rangle=\delta_{N,N'}\delta_{M,M'}\langle n,m|\hat{O}(r)|n',m'\rangle,
\end{equation}
where we use $|N, M\rangle (|n,m\rangle)$ to label the center-of-mass(relative) coordinate state. 

The only non-vanishing transformation coefficients are these between bases that satisfy the condition
\begin{equation}
\begin{aligned}
i+j&=N\\
k+l&=N-M\\
\epsilon_1+\epsilon_2&=N+n\\
m_1+m_2&=M+m,
\end{aligned}
\end{equation}
which is given by
\begin{equation}
\begin{aligned}
&<\epsilon_1m_1\epsilon_2m_2|NMnm>\\
=&\mathcal{C}\frac{\mathcal{A}_1\mathcal{A}_2}{\mathcal{A}_c\mathcal{A}_r}\sum_{i=0}^{\epsilon_1}\sum_{j=0}^{\epsilon_2}\sum_{k=0}^{\epsilon_1-m1}\sum_{l=0}^{\epsilon_2-m_2}(-1)^{2\epsilon_2-m_2-j-l}C_{\epsilon_1}^iC_{\epsilon_2}^jC_{\epsilon_1-m_1}^kC_{\epsilon_2-m_2}^l,
\end{aligned}
\end{equation}
where $\mathcal{C}=(1/\sqrt{2})^{2\epsilon_1+2\epsilon_2-m_1-m_2}$ and 
\begin{equation}
\mathcal{A}=\frac{(-1)^{min\{n,n-m\}}}{\sqrt{n!(n-m)!}},
\end{equation}
the subscripts of which are labels of the corresponding particle.
Derivation of the above transformation can be find in \ref{appendix_twobody}.

\subsection{Examples}

As a concrete example of implementing the procedure described above, a four-electron system is studied in this section. The table below lists the number of Slater determinants at various values of $K$s. The total angular momentum $M$ is fixed to be $-10$. 
\begin{center}\label{table}
  \begin{tabular}{ p{0.25\textwidth} || p{0.2\textwidth} | p{0.2\textwidth} |p{0.2\textwidth} }
    \hline
    K & 10 & 12 & 14 \\ \hline
    D1 & 5 & 41 & 144 \\ \hline
    D2 & 2 & 7 & 12 \\
    \hline
  \end{tabular}
\end{center}
$D1$ is the number of Slater determinants that may be part of the $\{K,M\}$ manifold before diagonalizing $Hcm$, $Lcm$ and $K^2$ operators (the process in Section \ref{sec_enumerating}). D2 is the number of Slater determinants after selecting the correct eigenvalue of $K$ and choosing the linear combinations that have the values $L_{CM}=0$ and $H_{CM}=1/2$ in the center-of-mass motion (the process in Section \ref{sec_diag} is performed).

For instance, there are 5 Slater determinant at $K=10$:
\begin{equation}
\begin{aligned}
&|\psi1\rangle=|0,\ 0\rangle|0,-1\rangle|0,-4\rangle|0,-5\rangle,\\
&|\psi2\rangle=|0,\ 0\rangle|0,-2\rangle|0,-3\rangle|0,-5\rangle,\\
&|\psi3\rangle=|0,-1\rangle|0,-2\rangle|0,-3\rangle|0,-4\rangle,\\
&|\psi4\rangle=|0,\ 0\rangle|0,-1\rangle|0,-2\rangle|0,-7\rangle,\\
&|\psi5\rangle=|0,\ 0\rangle|0,-1\rangle|0,-3\rangle|0,-6\rangle.
\end{aligned}
\end{equation}
The basis functions with $Lcm=0$, $Hcm=1/2$ and $K=10$ are the linear combinations of the above five Slater determinants,
\begin{equation}
\begin{aligned}
&|\Psi1\rangle=0.6727|\psi1\rangle-0.2212|\psi2\rangle+0.4946|\psi3\rangle+0.2760|\psi4\rangle-0.4216|\psi5\rangle,\\
&|\Psi2\rangle=-0.2805|\psi1\rangle-0.3335|\psi2\rangle+0.7458|\psi3\rangle-0.2760|\psi4\rangle+0.4216|\psi5\rangle.
\end{aligned}
\end{equation}

\begin{figure}
  \centering
  \includegraphics[angle=0,width=0.7\textwidth]{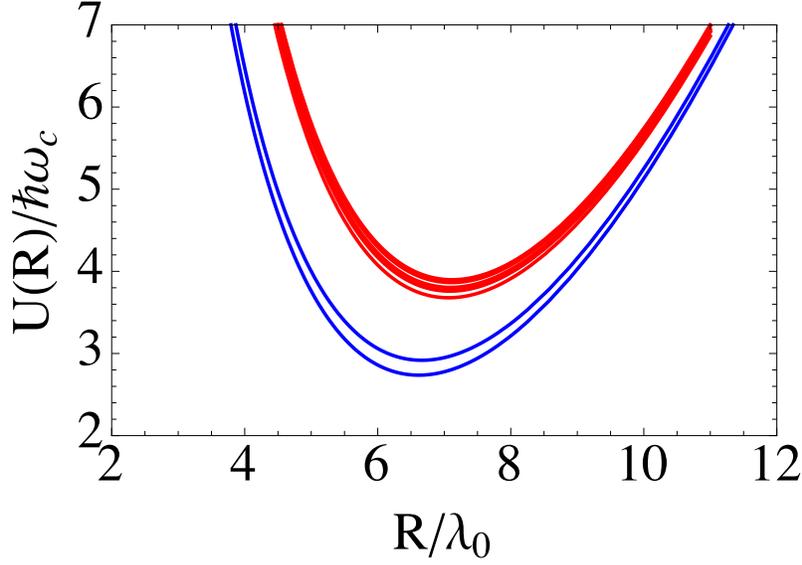} 
  \caption{Hyperradial potential curves. The blue lines (lower branch) are hyperradial potentials for $K=-M=10$, which correspond to the lowest Landau level. The red lines (upper branch) correspond to $K=-M+2=12$. The energy gap between separate clusters is approximately equal to the cyclotron excitation, while the smaller splittings within each branch are due to Coulomb interactions. The relative strength of the Coulomb interaction, $\kappa=\frac{e^2}{4\pi\epsilon\lambda_0}\frac{1}{\hbar\omega_c}$, is set to be 1, which is the typical order for experiments in gallium arsenide.}
  \label{fig_rcurve}
\end{figure}

The next step is to diagonalize the Coulomb interaction at fixed hyperradius, that is, the hyperradius is treated as an adiabatic parameter (as described is Section \ref{sec_CoulombMatrix}). The hyperradius potential curves are plotted in Fig.\ref{fig_rcurve}.

\begin{figure}
  \centering
  \includegraphics[angle=0,width=0.7\textwidth]{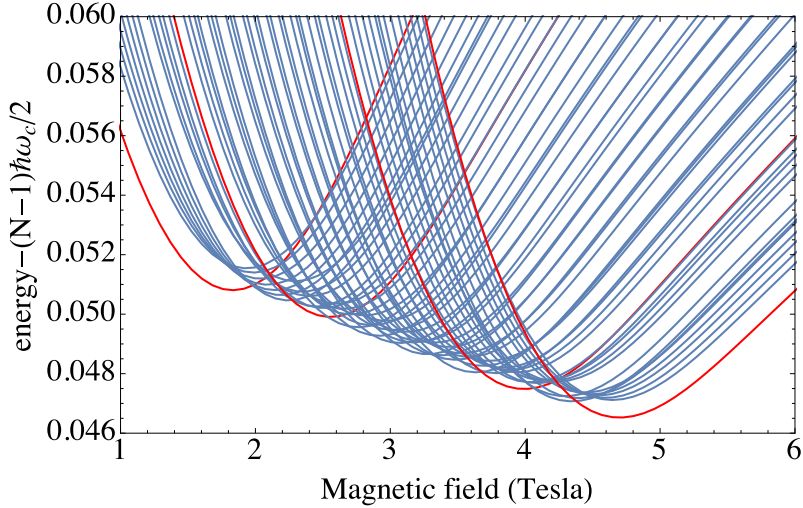} 
  \caption{The ground state energies of 8 particles in the lowest Landau level ($K=|M|$) at various $K$ values as a function of the magnetic field, using experimental parameters matching a cold gallium arsenide system\cite{footnote}. The magnetic field dependent zero-point energy has been subtracted for scale. The curves correspond to various $K$ values ranging from $K=28$ (filling factor $\nu=1$) to $K=84$ ($\nu=1/3$). Curves for quantum Hall states at $\nu=1$, $2/3$, $2/5$ and $1/3$ are marked as red lines from left to right respectively.}
  \label{fig_evsb}
\end{figure}

Fig.\ref{fig_evsb} shows an example of the calculation of a 8-particle system in the lowest Landau level ($K=|M|$). In the absence of coupling between different $K$ manifolds, the ground state energies are calculated by first diagonalizing the Coulomb interaction within a hyperangular $\{K,M\}$ manifold then solving the hyperradial Schr\"odinger's equation numerically with a restricted maximum hyperradius.

\section{Summary}\label{sec_summary}
In this work, we have established the representation of the $\{K, M\}$ manifold, the state space with fixed grand angular momentum quantum number $K$ and total relative angular momentum $M$, in terms of Slater determinants of single particle states. 
The Hilbert space spanned by the anti-symmetrized hyperspherical basis functions are the same as that spanned by the Slater determinants of single particle states. However, the grand angular momentum quantum number $K$ in the hyperspherical approach isolates a Hilbert subspace which does not have a natural analog in the independent particle picture. This subspace has been shown to be approximately separate from the hyperradial degree of freedom. Moreover, manifold $\{K, M\}$ involves single particle states in higher Landau levels, making it a nice framework for the study of the quantum Hall effect involving inter-Landau level physics\cite{Simon:2013,Sodemann:2013,Peterson:2013,Wooten:2013}, especially in the few-body limit. Formulas of the two-body interaction matrix elements in the Slater determinant basis are also derived. 

\section*{Acknowledgments}

This work has been supported in part by NSF grant PHY-1607180, and some computations have been carried out under the NSF XSEDE Resource Allocation No. TG-PHY150003.

\bibliography{draft}

\begin{thebibliography}{10}
\bibitem{footnote}
In typical systems, fractional quantum hall states appear at a higher magnetic
  field. different physical cut-offs of the system size correspond to changing
  of the particle density, which also changes the position of the fractional
  quantum hall states. in the present example, the maximum hyperradius cut-off
  is chosen such that the integer quantum hall effect appears at around 1
  tesla.
\end{thebibliography}

\appendix\label{appendix}

\section{Enumeration of Slater determinants of a fixed $\{K,M_{tot}\}$ manifold using contingency tables}\label{appendixKM}

As stated earlier, listing the complete set of $N$-particle Slater determinants spanning a fixed $\{K,M_{tot}\}$ manifold is equivalent to finding the complete list of sets of single-particle orbitals that satisfy the rules listed in Section \ref{sec_enumerating}.  Because $N$, $K$, $M_{tot}$, and all of the single-particle orbital quantum numbers are integers, we can use techniques from number theory to find the allowed sets of single-particle orbitals.  One efficient method for finding the Slater determinants is find the integer partitions on $K$ and on the number of total excitations in the desired system, and use those partitions to construct contingency tables. 

We start by noting that the grand angular momentum $K$ is the total order of the harmonic polynomial part of the final, $N$-particle Slater determinants.  
Since $K$ is the order of the polynomial we seek,
we must include all Slater determinants in which
the polynomial orders of the individual orbitals sum up to $K$.
The order of the polynomial
of the single-particle orbital $|\epsilon,m\rangle$ is $2\epsilon - m$,
thus the restriction is
\begin{align}
  \label{eq_Krestriction}
  K = \sum_{j=1}^N \left(2\epsilon_j - m_j\right) = 2n_{\epsilon} - M_{tot}.
\end{align}
Here, $n_{\epsilon}$ is the total number of Landau level excitations.
This restriction is equivalent to finding the integer partitions of $K$ of length less than or equal to $N$, where the partition of an integer $A$ of length $B$ is simply the list of all possible (ordered) sets of integers that sum up to $A$ with exactly $B$ elements in each ordered set.
For example,
for a three particle system and $K=4$, the allowed partitions are
$[4,0,0]$, $[3,1,0]$, $[2,2,0]$, and $[2,1,1]$. 
For this system, there are 4 different ways to divide the total polynomial order, $K$, among the 3 single-particle orbitals without any reference to the information about the specific orbitals chosen.

\begin{figure}
  \centering
  \includegraphics[angle=0,width=0.75\textwidth]{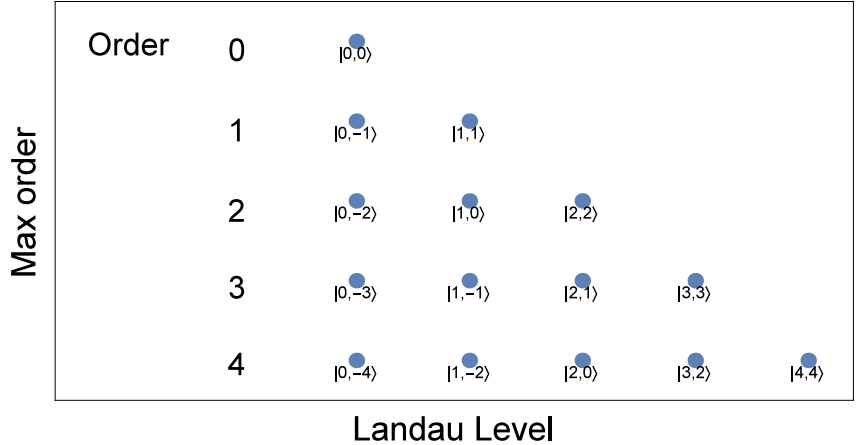} 
  \caption{(Color online) 
    Diagram describing the ordering of the single-particle
    orbitals $|\epsilon,m\rangle$ by Landau level and order of the
    polynomial.
  }
  \label{fig_contingency}
\end{figure}
Figure \ref{fig_contingency} shows how the single-particle orbitals are arranged by
the order of their polynomial part
and Landau level.
Notice that the integer partitions of $K$ can be
recast into an alternate notation which identifies the number of orbitals to pick from each row of
Fig.~\ref{fig_contingency}.
Taking partition $[4,0,0]$, for example,
it is recast to $\{2,0,0,0,1\}$,
which means we can pick two orbitals of order 0,
no orbitals of order 1,
no orbitals of order 2,
no orbitals of order 3, and
one orbital of order 4.
The rest of the partitions in this example become
$\{1,1,0,1\}$, $\{1,0,2\}$, and $\{0,2,1\}$, respectively.
The second observation is that there is only one orbital
with order 0 polynomial ($|0,0 \rangle$),
only two orbitals of order 1 ($|0,-1\rangle$ and $|1,1\rangle$),
three orbitals of order 2 ($|0,-2\rangle$, $|1,0\rangle$, and $|2,2\rangle$),
and so on.
This presents an additional restriction;
namely,
in this example the $\{2,0,0,0,1\}$ set of column choices is not allowed.

We perform a similar partitioning and restriction by fixing $M$.
This can be done by brute force
to enumerate every possibility given the restrictions on $K$,
but this is inefficient.
Alternatively,
the restriction on $M$ can be restated as a restriction on the number
of Landau level excitations $n_{\epsilon}$.
Looking back at Eq.~\eqref{eq_Krestriction},
the total number of Landau level excitations is equal to $n_{\epsilon}=(K+M_{tot})/2$.
Dividing the number of excitations among the different Landau
levels is again equivalent to finding the integer partitions,
this time of $n_{\epsilon}$, restricted by the number of particles.
Continuing the example for $N=3$ and $K=4$,
if $M_{tot}=0$, then this implies that $n_{\epsilon}=2$
and its integer partitions are $[2,0,0]$ or $[1,1,0]$.
Recasting these in terms of the number of orbitals to pick
from each column of Fig.~\ref{fig_contingency},
yields $\{2,0,1\}$ and $\{1,2\}$, respectively.

Combining the results, from the restrictions on $K$ we have all possible lists of the number of orbitals to choose from each row of Figure \ref{fig_contingency}, and from the restrictions on $n_\epsilon$ we have all possible lists of the number of orbitals to choose from each column of Figure \ref{fig_contingency}.
Said differently,
the number of Slater determinants that contribute to the subspace
of fixed $M$ and $K$ is equivalent to counting
all possible triangular binary rectangular contingency tables
with specified margins.
The row margins, that is, the restricted sums over row elements,
are given by the set of numbers that count how many times a given number
appears in the integer partitions of $K$.
The column margins, that is, the restricted sums over column elements,
are given by the set of numbers that count how many times a given number
appears in the integer partitions of $n_{\epsilon}$.
Both integer partitions are restricted to have no more than $N$ parts.
There are known algorithms to efficiently enumerate the possibilities,
but the problem scales polynomially with $K$ and $n_{\epsilon}$.

Continuing the example,
there are six possible contingency matrices:
\begin{align}
  \begin{tabular}{c|ccc}
    \multicolumn{1}{c}{}
      & 2 & 0 & 1 \\ \cline{2-4}
    1 & . & 0 & 0 \\
    1 & . & . & 0 \\
    0 & . & . & . \\
    1 & . & . & . 
  \end{tabular}
  \qquad 
  \begin{tabular}{c|cc}
    \multicolumn{1}{c}{}
      & 1 & 2 \\ \cline{2-3}
    1 & . & 0 \\
    1 & . & . \\
    0 & . & . \\
    1 & . & .
  \end{tabular}
  \qquad
  \begin{tabular}{c|ccc}
    \multicolumn{1}{c}{}
      & 2 & 0 & 1 \\ \cline{2-4}
    1 & . & 0 & 0 \\
    0 & . & . & 0 \\
    2 & . & . & . 
  \end{tabular}    \\ \nonumber \\
  \begin{tabular}{c|cc}
    \multicolumn{1}{c}{}
      & 1 & 2 \\ \cline{2-3}
    1 & . & 0 \\
    0 & . & . \\
    2 & . & . 
  \end{tabular}
  \qquad
  \begin{tabular}{c|ccc}
    \multicolumn{1}{c}{}
      & 2 & 0 & 1 \\ \cline{2-4}
    0 & . & 0 & 0 \\
    2 & . & . & 0 \\
    1 & . & . & . 
  \end{tabular}
  \qquad 
  \begin{tabular}{c|cc}
    \multicolumn{1}{c}{}
      & 1 & 2 \\ \cline{2-3}
    0 & . & 0 \\
    2 & . & . \\
    1 & . & . 
  \end{tabular}.
\end{align}
The row and column restrictions are indicated to the left of and above
the matrices, respectively.
The upper triangle of each is made up of zeros as there are no
orbitals to pick from in these locations.
Dots indicate elements that are unknown within the matrix
(not including trivial zero rows or zero columns).
By inspection, it is straightforward to enumerate all possible matrices,
namely
\begin{align}
  \begin{tabular}{c|ccc}
    \multicolumn{1}{c}{}
      & 2 & 0 & 1 \\ \cline{2-4}
    1 & 1 & 0 & 0 \\
    1 & 1 & 0 & 0 \\
    0 & 0 & 0 & 0 \\
    1 & 0 & 0 & 1 
  \end{tabular}
  \qquad
  \begin{tabular}{c|cc}
    \multicolumn{1}{c}{}
      & 1 & 2 \\ \cline{2-3}
    1 & 1 & 0 \\
    1 & 0 & 1 \\
    0 & 0 & 0 \\
    1 & 0 & 1
  \end{tabular}
  \qquad
  \begin{tabular}{c|ccc}
    \multicolumn{1}{c}{}
      & 2 & 0 & 1 \\ \cline{2-4}
    1 & 1 & 0 & 0 \\
    0 & 0 & 0 & 0 \\
    2 & 1 & 0 & 1 
  \end{tabular}
  \qquad 
  \begin{tabular}{c|cc}
    \multicolumn{1}{c}{}
      & 1 & 2 \\ \cline{2-3}
    0 & 0 & 0 \\
    2 & 1 & 1 \\
    1 & 0 & 1
  \end{tabular}
\end{align}
which translate to the Slater determinants
$|0,0\rangle|0,-1\rangle|2,1\rangle$,
$|0,0\rangle|1,1\rangle|1,-1\rangle$,
$|0,0\rangle|0,-2\rangle|2,2\rangle$, and
$|0,-1\rangle|1,1\rangle|1,0\rangle$, respectively.
It is observable that by restricting $n_{\epsilon}$,
all of these Slater determinants do in fact have $M_{tot}=0$,
as was desired.
In practice,
we enumerate the possible configurations
of a single contingency matrix by starting with the first column.
For the first column
there are $\binom{N}{c_1}$ possible 0-1 column vectors,
where $c_1$ is the first column restriction.
Choosing one of the allowed 0-1 column vectors,
the resulting counting problem is identical to the previous step,
though with a new matrix of one less column and new row restrictions.

\section{Center-of-mass operators}\label{appendix_operator}
Expressed in terms of the ladder operators,
the single-particle coordinates are
\begin{align}
  x &= \frac{1}{2^{1/2}}\left(a + a^{\dagger} + b + b^{\dagger} \right) \qquad
  &y= \frac{\imath}{2^{1/2}}\left(a - a^{\dagger} -b + b^{\dagger}  \right) \\
  \frac{\partial}{\partial x} &=
  \frac{1}{2^{3/2}} \left(a - a^{\dagger} + b - b^{\dagger} \right) \qquad
  &\frac{\partial}{\partial y} =
  \frac{-\imath}{2^{3/2}}\left(-a - a^{\dagger} + b + b^{\dagger} \right).
\end{align}
The center of mass coordinates and its partial derivatives,
expressed in the single-particle coordinates, are
\begin{align}
  x_{CM} &= \frac{1}{N} \sum_{j=1}^N x_j \qquad
  & y_{CM} = \frac{1}{N} \sum_{j=1}^N y_j \\
 \frac{\partial}{\partial x_{CM}} &=
    \sum_{j=1}^N \frac{\partial}{\partial x_j} \qquad 
  &\frac{\partial}{\partial y_{CM}} =
    \sum_{j=1}^N \frac{\partial}{\partial y_j}.
\end{align}
In the following, it is also useful to use the commutator relations,
which are
\begin{align}
  \label{eq_commutatora}
  \left[ a, a^{\dagger} \right] &= 1 \\
  \label{eq_commutatorb}
  \left[ b, b^{\dagger} \right] &= 1 \\
  \label{eq_commutatorab}
  \left[ a, b \right] &= \left[ a^{\dagger} , b^{\dagger} \right]
  = \left[ a^{\dagger}, b \right] = \left[ a, b^{\dagger} \right] =0.
\end{align}

\subsection{Angular momentum}
The angular momentum of the center of mass becomes
\begin{align}
  \label{eq_L}
  L_{CM}
  & = -\imath \hbar \left( x_{CM} \frac{\partial}{\partial y_{CM}} -
  y_{CM} \frac{\partial}{\partial x_{CM}} \right) \nonumber \\
  & = -\frac{\imath \hbar}{N} \sum_{j=1}^N \sum_{k=1}^N
  \left( x_j \frac{\partial}{\partial y_k}
  - y_j \frac{\partial}{\partial x_k} \right) \nonumber \\
  & = -\frac{\hbar}{4N} \sum_{j=1}^N \sum_{k=1}^N
  \bigg[
    \left(a_j + a_j^{\dagger} + b_j + b_j^{\dagger} \right)
    \left(-a_k - a_k^{\dagger} + b_k + b_k^{\dagger} \right) \nonumber \\
    &\qquad \qquad \qquad \quad +
    \left(a_j - a_j^{\dagger} - b_j + b_j^{\dagger} \right)
    \left(a_k - a_k^{\dagger} + b_k - b_k^{\dagger} \right)
    \bigg] \nonumber \\
  & = -\frac{\hbar}{2N} \sum_{j=1}^N \sum_{k=1}^N
  \left[ a_jb_k - b_ja_k + a_j^{\dagger}b_k^{\dagger} - b_j^{\dagger}a_k^{\dagger}
    - a_ja_k^{\dagger} - a_j^{\dagger}a_k + b_jb_k^{\dagger} + b_j^{\dagger}b_k  \right]
  \nonumber \\
  & = -\frac{\hbar}{2N} \sum_{j=1}^N \sum_{k=1}^N
  \left[-a_ja_k^{\dagger} -a_j^{\dagger}a_k +b_jb_k^{\dagger} +b_j^{\dagger}b_k  \right].
\end{align}
It can be seen that the $ab$ terms cancel ($a$ and $b$ commute)
when summing over all indices.

\subsection{Center of mass Hamiltonian}
We wish to transform the center of mass Hamiltonian $H_{CM}$
in terms of the raising and lowering operators
of the quantum Hall problem.
In cyclotron units and assuming equal mass particles,
the center of mass Hamiltonian is
\begin{align}
  \label{eq_Hcm}
  H_{CM} = -\frac{1}{2 N} \left(
  \frac{\partial^2}{\partial x^2_{CM}}
  + \frac{\partial^2}{\partial y^2_{CM}} \right)
  + \frac{N}{8} \left( x^2_{CM} + y^2_{CM} \right) + \frac{1}{2\hbar} L_{CM}
\end{align}
The $x^2_{CM}+y^2_{CM}$ term yields
\begin{align}
  \label{eq_xycm}
  x^2_{CM} +y^2_{CM} & = \frac{1}{N^2} \sum_{j=1}^N \sum_{k=1}^N
  \left(x_jx_k + y_jy_k\right) \nonumber \\ 
  &= \frac{1}{2 N^2} \sum_{j=1}^N \sum_{k=1}^N
  \bigg[
    \left(a_j + a_j^{\dagger} + b_j + b_j^{\dagger} \right)
    \left(a_k + a_k^{\dagger} + b_k + b_k^{\dagger} \right) \nonumber \\
    &\qquad \qquad \quad \quad-
    \left(a_j - a_j^{\dagger} - b_j + b_j^{\dagger} \right)
    \left(a_k - a_k^{\dagger} - b_k + b_k^{\dagger} \right)
    \bigg] \nonumber \\
  &= \frac{1}{N^2} \sum_{j=1}^N \sum_{k=1}^N
  \left[a_jb_k + b_ja_k + a_j^{\dagger}b_k^{\dagger} + b_j^{\dagger}a_k^{\dagger}
    + a_ja_k^{\dagger} + a_j^{\dagger}a_k + b_jb_k^{\dagger} + b_j^{\dagger}b_k  \right]
\end{align}
The $\frac{\partial^2}{\partial x^2_{CM}}+\frac{\partial^2}{\partial y^2_{CM}}$
terms become 
\begin{align}
  \label{eq_dxycm}
  \frac{\partial^2}{\partial x^2_{CM}} + \frac{\partial^2}{\partial y^2_{CM}}
  &= \sum_{j=1}^N \sum_{k=1}^N
  \left(
  \frac{\partial}{\partial x_j} \frac{\partial}{\partial x_k} +
  \frac{\partial}{\partial y_j} \frac{\partial}{\partial y_k}
  \right) \nonumber \\
  &= \frac{1}{8} \sum_{j=1}^N \sum_{k=1}^N
  \bigg[
    \left(a_j - a_j^{\dagger} + b_j - b_j^{\dagger} \right)
    \left(a_k - a_k^{\dagger} + b_k - b_k^{\dagger} \right) \nonumber \\
    &\qquad \quad \quad \quad-
    \left(-a_j - a_j^{\dagger} + b_j + b_j^{\dagger} \right)
    \left(-a_k - a_k^{\dagger} + b_k + b_k^{\dagger} \right)
    \bigg] \nonumber \\
  &= \frac{1}{4} \sum_{j=1}^N \sum_{k=1}^N
  \left(
  a_jb_k + b_ja_k + a_j^{\dagger}b_k^{\dagger} + b_j^{\dagger}a_k^{\dagger}
  - a_ja_k^{\dagger} - a_j^{\dagger}a_k - b_jb_k^{\dagger} - b_j^{\dagger}b_k 
  \right)
\end{align}
Combining  Eq.~\eqref{eq_L}, Eq.~\eqref{eq_xycm}, and Eq.~\eqref{eq_dxycm}
as in the center of mass Hamiltonian, Eq.~\eqref{eq_Hcm}, yields
\begin{align}
  H_{CM} =
  \frac{1}{2N} \sum_{j=1}^N \sum_{k=1}^N
  \left( a_ja_k^{\dagger} + a_j^{\dagger}a_k  \right)
\end{align}

\subsection{Hyperangular operator}
The squared grand angular momentum operator $\hat{K}^2$
in 2-D takes the form~\cite{avery}
\begin{align}
  \hat{K}^2 =& -R^2 \nabla^2 + \sum_{j=1}^N\sum_{k=1}^N
  \left(
    x_j x_k \frac{\partial}{\partial x_j}\frac{\partial}{\partial x_k}
  + x_j y_k \frac{\partial}{\partial x_j}\frac{\partial}{\partial y_k}
  + y_j x_k \frac{\partial}{\partial y_j}\frac{\partial}{\partial x_k}
  + y_j y_k \frac{\partial}{\partial y_j}\frac{\partial}{\partial y_k}\right)
  \nonumber \\
  &+(2N-1)\sum_{j=1}^N\left(
  x_j\frac{\partial}{\partial x_j} + y_j\frac{\partial}{\partial y_j}
  \right)
\end{align}
where $R$ is the hyperadius and $\nabla^2$ is the Laplacian operator.
Expanding out these terms yields
\begin{align}
  \label{eq_k2}
  \hat{K}^2 =&  \sum_{j=1}^N\sum_{k=1}^N
  \left(
    x_j x_k \frac{\partial}{\partial x_j}\frac{\partial}{\partial x_k}
  + x_j y_k \frac{\partial}{\partial x_j}\frac{\partial}{\partial y_k}
  + y_j x_k \frac{\partial}{\partial y_j}\frac{\partial}{\partial x_k}
  + y_j y_k \frac{\partial}{\partial y_j}\frac{\partial}{\partial y_k}\right)
  \nonumber \\
  &-\sum_{j=1}^N\sum_{k=1}^N\left( x_j^2+y_j^2 \right)
  \left( \frac{\partial^2}{\partial x_k^2}
  + \frac{\partial^2}{\partial y_k^2} \right) 
  +(2N-1)\sum_{j=1}^N\left(
  x_j\frac{\partial}{\partial x_j} + y_j\frac{\partial}{\partial y_j}
  \right).
\end{align}

\subsubsection{Diagonal terms}
For ease of computation,
we break up our calculation of $\hat{K}^2$ into its diagonal ($j=k$)
and off-diagonal ($j \ne k$) parts.
There are two contributions to the diagonal terms;
the $j=k$ terms of the double sum
and the single summation of Eq.~\eqref{eq_k2}.
First,
leaving off the $j$ label for convenience,
the double sum becomes
\begin{align}
  \frac{1}{4} \sum_{j=k}
  \bigg( &
  aaa^{\dagger}a^{\dagger} +aa^{\dagger}aa^{\dagger}
  + a^{\dagger}aa^{\dagger}a + a^{\dagger}a^{\dagger}aa
  +bbb^{\dagger}b^{\dagger} + bb^{\dagger}bb^{\dagger}
  +b^{\dagger}bb^{\dagger}b+b^{\dagger}b^{\dagger}bb
  \nonumber \\ &
  -2\left[
    aa^{\dagger}bb^{\dagger}+aa^{\dagger}b^{\dagger}b
    +a^{\dagger}abb^{\dagger}+a^{\dagger}ab^{\dagger}b
    \right]
  \nonumber \\ &
  +a^{\dagger}bb^{\dagger}b^{\dagger} - a^{\dagger}b^{\dagger}b^{\dagger}b
  +a^{\dagger}aab-aaa^{\dagger}b
  -abbb^{\dagger}+ab^{\dagger}bb
  +aa^{\dagger}a^{\dagger}b^{\dagger}-a^{\dagger}a^{\dagger}ab^{\dagger}
  \bigg).
\end{align}
Applying the commutator rules
Eqs.~\eqref{eq_commutatora} and~\eqref{eq_commutatorb}
to put as many of the terms into
number operator form (e.g. $a^{\dagger}a$) as possible,
after much algebra yields
\begin{align}
  \label{eq_diagonal1}
  \sum_{j=k}
  \left( 1+a^{\dagger}aa^{\dagger}a +b^{\dagger}bb^{\dagger}b-2a^{\dagger}ab^{\dagger}b
  +a^{\dagger}b^{\dagger} -ab \right).
\end{align}
Second,
the single summation of Eq.~\eqref{eq_k2} becomes
\begin{align}
  \label{eq_diagonal2}
  (2N-1)\sum_{j=1}^N\left(
  x_j\frac{\partial}{\partial x_j} + y_j\frac{\partial}{\partial y_j}
  \right)
  = (2N-1)\sum_{j=1}^N\left(-1 + ab - a^{\dagger}b^{\dagger} \right)
\end{align}
Combining Eqs.~\eqref{eq_diagonal1} and~\eqref{eq_diagonal2} yields
\begin{align}
  \label{eq_diagonalall}
  -2N(N-1) + \sum_{j=1}^N
  \left(
  \left[a_j^{\dagger}a_j-b_j^{\dagger}b_j\right]^2
  +2(N-1) \left[ a_jb_j - a_j^{\dagger}b_j^{\dagger}\right]
  \right)
\end{align}
where additional factors of $N$ have come from pulling out the $1$'s
from the summations.

\subsubsection{Off-diagonal terms}
The off-diagonal parts of the $\hat{K}^2$ operator come
from the double summation,
where it is assumed that $j \ne k$.
There are a total of $N(N-1)$ terms in this summation.
In ladder operator form,
putting all $j$ indices to the left and all $k$ indices to the right,
the off-diagonal part is
\begin{align}
  \frac{1}{2} \sum_j\sum_{k \ne j}
  \left(
  \left[aa^{\dagger}+2a^{\dagger}b^{\dagger}+bb^{\dagger} \right]_j
  \left[aa^{\dagger}-2ab+bb^{\dagger} \right]_k
  +\left[a^{\dagger}a + 2ab +b^{\dagger}b \right]_j
  \left[a^{\dagger}a-2a^{\dagger}b^{\dagger}+b^{\dagger}b \right]_k
  \right)
\end{align}
where we use a compact notation where a subscript on the parenthesis
implies all elements within the parenthesis have that label.
Applying the commutator rules
Eqs.~\eqref{eq_commutatora} and~\eqref{eq_commutatorb} yields
\begin{align}
  \frac{1}{2} \sum_j\sum_{k \ne j}
  &\bigg( 4
  + 2\left[a^{\dagger}a-2ab+b^{\dagger}b \right]_k
  + 2\left[a^{\dagger}a+2a^{\dagger}b^{\dagger}+b^{\dagger}b \right]_j
  \nonumber \\ &
  +\left[a^{\dagger}a+2a^{\dagger}b^{\dagger}+b^{\dagger}b \right]_j
  \left[a^{\dagger}a-2ab+b^{\dagger}b \right]_k
  +\left[a^{\dagger}a + 2ab +b^{\dagger}b \right]_j
  \left[a^{\dagger}a-2a^{\dagger}b^{\dagger}+b^{\dagger}b \right]_k
  \bigg).
\end{align}
Expanding and simplifying yields
\begin{align}
  \label{eq_offdiagonalall}
  &2N(N-1)
  + 2(N-1)\sum_{j=1}^N
  \left(
  a_j^{\dagger}a_j +b_j^{\dagger}b_j-a_jb_j+a_j^{\dagger}b_j^{\dagger}
  \right)
  \nonumber \\ &
  + \sum_j\sum_{k \ne j}
  \left(
  \left[a^{\dagger}a+b^{\dagger}b \right]_j\left[a^{\dagger}a+b^{\dagger}b\right]_k
  -2\left(a_j^{\dagger}b_j^{\dagger}a_kb_k+a_jb_ja_k^{\dagger}b_k^{\dagger} \right)
  \right).
\end{align}

\subsection{Full $\hat{K}^2$ operator}
Combining Eqs.~\eqref{eq_diagonalall} and~\eqref{eq_offdiagonalall} yields
\begin{align}
  \hat{K}^2 = & \sum_{j=1}^N
  \left(
  \left[a^{\dagger}a - b^{\dagger}b\right]_j^2
  +(2N-2) \left[a^{\dagger}a+b^{\dagger}b\right]_j
  \right)
  \nonumber \\ &
  +\sum_j \sum_{k \ne j}
  \left(
  \left[a^{\dagger}a+b^{\dagger}b \right]_j\left[a^{\dagger}a+b^{\dagger}b\right]_k
  -2\left(a_j^{\dagger}b_j^{\dagger}a_kb_k+a_jb_ja_k^{\dagger}b_k^{\dagger} \right)
  \right).
\end{align}
Note that only the last term can lead to coupling between Slater determinants.

\section{Transformation of the two-body basis}\label{appendix_twobody}
A single particle state $|\epsilon,m\rangle$ can be generated with raising operators from the vacuum state $|0>$:

\begin{equation}
|\epsilon,m>=\mathcal{A}\hat{a}^\dagger\hat{b}^{\dagger \epsilon-m}|0>,
\end{equation}
or be eliminated using lowering operators:
\begin{equation}\label{lower}
\mathcal{A}\hat{a}\hat{b}^{\epsilon-m}|\epsilon,m>=|0>,
\end{equation}
where $\mathcal{A}$ is the normalization factor
\begin{equation}\label{a}
\mathcal{A}=\frac{(-1)^{min\{\epsilon,\epsilon-m\}}}{\sqrt{\epsilon!(\epsilon-m)!}},
\end{equation}
and
\begin{equation}
min\{\epsilon,n-m\}=\epsilon-\frac{m+|m|}{2}.
\end{equation}
The center-of-mass (relative coordinate) state $|N,M\rangle$ ($|n,m\rangle$) can be expressed as the above form as well, using the corresponding ladder operators:

\begin{align}
  a_{CM} = \frac{1}{\sqrt{2}}\left(a_1+a_2\right), \\
  a_{r} = \frac{1}{\sqrt{2}}\left(a_1-a_2\right),  \\
  b_{CM} = \frac{1}{\sqrt{2}}\left(b_1+b_2\right),  \\
  a_{r} = \frac{1}{\sqrt{2}}\left(b_1-b_2\right).
\end{align}
Using Eq. (\ref{lower}), $|\epsilon_1m_1\epsilon_2m_2>$ can be written as
\begin{equation}
<\epsilon_1m_1\epsilon_2m_2|=<0|\mathcal{A}_1\mathcal{A}_2\hat{a}_1^{\epsilon_1}\hat{b}_1^{\epsilon_1-m_1}\hat{a}_2^{\epsilon_2}\hat{b}_2^{\epsilon_2-m_2}.
\end{equation}
The single particle basis Ladder operators can be expanded with the CM-relative coordinate basis operators as
\begin{equation}
\begin{aligned}
&\hat{a}_1^{\epsilon_1}\hat{b}_1^{\epsilon_1-m_1}\hat{a}_2^{\epsilon_2}\hat{b}_2^{\epsilon_2-m_2}\\
=&\mathcal{C}(\hat{a}_c+\hat{a}_r)^{\epsilon_1}(\hat{b}_c+\hat{b}_r)^{\epsilon_1-m1}(\hat{a}_c-\hat{a}_r)^{\epsilon_2}(\hat{b}_c-\hat{b}_r)^{\epsilon_2-m_2}\\
=&\mathcal{C}\sum_{i=0}^{\epsilon_1}\sum_{j=0}^{\epsilon_2}\sum_{k=0}^{\epsilon_1-m_1}\sum_{l=0}^{\epsilon_2-m_2}(-1)^{2\epsilon_2-m_2-j-l}\\
&\ \ \times C_{\epsilon_1}^iC_{\epsilon_2}^jC_{\epsilon_1-m_1}^kC_{\epsilon_2-m_2}^l\hat{a}_c^{i+j}\hat{a}_r^{\epsilon_1+\epsilon_2-i-j}\hat{b}_c^{k+l}\hat{b}_r^{\epsilon_1+\epsilon_2-m_1-m_2-k-l}
\end{aligned}
\end{equation}
where $\mathcal{C}=(1/\sqrt{2})^{2\epsilon_1+2\epsilon_2-m_1-m_2}$. The only survived terms in the inner product in the above summation are those that satisfy
\begin{equation}
\begin{aligned}
i+j&=N\\
k+l&=N-M\\
\epsilon_1+\epsilon_2&=N+n\\
m_1+m_2&=M+m,
\end{aligned}
\end{equation}
leaving the final transformation coefficient as
\begin{equation}
\begin{aligned}
&<\epsilon_1m_1\epsilon_2m_2|NMnm>\\
=&\mathcal{C}\frac{\mathcal{A}_1\mathcal{A}_2}{\mathcal{A}_c\mathcal{A}_r}\sum_{i=0}^{\epsilon_1}\sum_{j=0}^{\epsilon_2}\sum_{k=0}^{\epsilon_1-m1}\sum_{l=0}^{\epsilon_2-m_2}(-1)^{2\epsilon_2-m_2-j-l}C_{\epsilon_1}^iC_{\epsilon_2}^jC_{\epsilon_1-m_1}^kC_{\epsilon_2-m_2}^l,
\end{aligned}
\end{equation}where factor $\mathcal{A}$s are given by Eq. (\ref{a}).

\end{document}